\documentclass[12pt]{article}
\usepackage{hyperref}
\usepackage[english]{babel}
\usepackage{amsmath}
\usepackage{latexsym}
\usepackage{amssymb}
\usepackage[all]{xy}
\usepackage{epsfig}
\usepackage{psfrag}

\numberwithin{equation}{section} \numberwithin{table}{section}
\numberwithin{figure}{section}

\begin{document}



\begin{titlepage}
   \begin{flushright}
{\small MPP-2014-126}
  \end{flushright}

   \begin{center}

     \vspace{20mm}

     {\LARGE \bf On holographic entanglement entropy of non-local field theories

     \vspace{3mm}}

     \vspace{10mm}

    Da-Wei Pang

     \vspace{5mm}

      {\small \sl Max-Planck-Institut f\"{u}r Physik (Werner-Heisenberg-Institut)\\
      F\"{o}hringer Ring 6, 80805 M\"{u}nchen, Germany}\\

     {\small \tt dwpang@mppmu.mpg.de}
     \vspace{10mm}

   \end{center}

\begin{abstract}
\baselineskip=18pt
We study holographic entanglement entropy of non-local field theories both at extremality and finite temperature. The gravity duals, constructed in arXiv:1208.3469 [hep-th], are characterized by a parameter $w$. Both the zero temperature backgrounds and the finite temperature counterparts are exact solutions of Einstein-Maxwell-dilaton theory. For the extremal case we consider the examples with the entangling regions being a strip and a sphere. We find that the leading order behavior of the entanglement entropy always exhibits a volume law when the size of the entangling region is sufficiently small. We also clarify the condition under which the next-to-leading order result is universal. For the finite temperature case we obtain the analytic expressions both in the high temperature limit and in the low temperature limit. In the former case the leading order result approaches the thermal entropy, while the finite contribution to the entanglement entropy at extremality can be extracted by taking the zero temperature limit in the latter case. Moreover, we observe some peculiar properties of the holographic entanglement entropy when $w=1$.
\end{abstract}
\setcounter{page}{0}
\end{titlepage}

\pagestyle{plain} \baselineskip=19pt

\tableofcontents

\section{Introduction}
The entanglement entropy plays a very important role in quantum many-body physics,
as it provides a universal measure of the degrees of freedom for an arbitrary subsystem.
This enables us to extract many underlying features of the system under investigation.
It has been well known that for a local quantum field theory with a UV fixed point,
the entanglement entropy obeys the so-called ¡®area law¡¯~\cite{Eisert:2008ur}. The area law states that
for a given subsystem $A$, the entanglement entropy contains a UV divergent part, whose
coefficient is proportional to the area of the boundary $\partial A$ of the subsystem. Intuitively,
the area law follows from the locality of the corresponding quantum field theory: The
interactions are very short-range and the amount of entanglement could be significantly
reduced.

It is also well known that there are many examples in quantum many-body systems whose
entanglement entropy obeys a ¡®volume law¡¯. For example, generic excited states~\cite{Page:1993df} and a
spin system with non-local random interactions between any two pairs of spins~\cite{Vitagliano:2010db}. One
may expect that the entanglement entropy obeys a volume law if proper non-local field
theory is considered. It should be emphasized that not all non-local field theories will
lead to volume law for the entanglement entropy. A counter example was given in~\cite{Nezhadhaghighi:2013mba}.

The AdS/CFT correspondence~\cite{Maldacena:1997re, Aharony:1999ti} asserts the equivalence between a weakly coupled
gravity theory in Anti-de Sitter (AdS) spacetime and a strongly coupled conformal field
theory (CFT) living on the boundary of that AdS spacetime, hence it provides powerful
tools for analysing the dynamics of strongly coupled field theories. Based on AdS/CFT,
Ryu and Takayanagi proposed the following holographic formula for evaluating the entanglement
entropy of a given subsystem $A$ in the large N limit~\cite{Ryu:2006bv, Ryu:2006ef},
\begin{equation}
S_{A}=\frac{\rm Area(\gamma_{A})}{4G_{N}}.
\end{equation}
Here $\gamma_{A}$ denotes the minimal area surface whose boundary coincides with the boundary
of $A$, $\partial A=\partial\gamma_{A}$ and $G_{N}$ is the Newton constant. Moreover, it has been confirmed in~\cite{Ryu:2006bv, Ryu:2006ef}
that the area law holds for strongly coupled CFTs with AdS duals. For reviews on this
fascinating topic, see~\cite{Nishioka:2009un, Takayanagi:2012kg}.

Since there have been enormous examples where the entanglement entropy exhibits a
volume law, it would be interesting to see if such a volume law can be realized holographically.
Previous holographic examples on this volume-law behavior include a class
of non-local field theories~\cite{Barbon:2008sr, Barbon:2008ut}, flat spacetime~\cite{Li:2010dr} and non-commutative super Yang-
Mills theory~\cite{Fischler:2013gsa, Karczmarek:2013xxa}. Recently the volume law of holographic entanglement entropy for
a simple class of non-local field theories was confirmed both analytically and numerically
in~\cite{Shiba:2013jja}, where precise agreement with holographic calculations was also observed. For the latest developments on the volume-law behavior of holographic entanglement entropy, see~\cite{Mollabashi:2014qfa, Kol:2014nqa}.

The gravity dual of the non-local field theory considered in~\cite{Shiba:2013jja} was obtained by a formula
proposed in~\cite{Nozaki:2012zj}, based on the continuum limit of MERA (cMERA). Here MERA stands
for multi-scale entanglement renormalization ansatz, which was conjectured in~\cite{Swingle:2009bg} to
be equivalent to AdS/CFT. This conjecture enables us to relate MERA with gravity.
Consider the following $(d+2)-$dimensional gravity dual
\begin{equation}
ds^{2}=g_{uu}du^{2}+\frac{e^{2u}}{\epsilon^{2}}\sum\limits^{d}_{i=1}dx_{i}^{2}+g_{tt}dt^{2},
\end{equation}
where $\epsilon$ denotes the UV cutoff and $u$ is the radial coordinate. The boundary of this
spacetime is located at $u = 0$. If we consider the non-local Hamiltonian of the following form
\begin{equation}
H=\int d^{d}x[\frac{1}{2}(\partial_{t}\phi)^{2}+B_{0}\phi e^{A_{0}(-\partial_{i}\partial_{i})^{w/2}}\phi],
\end{equation}
where $A_{0}, B_{0}$ are positive constants, the corresponding gravity dual is given by~\cite{Nozaki:2012zj}
\begin{equation}
\label{nlgsec1}
ds^{2}\propto A_{0}^{2}\frac{dz^{2}}{z^{2(w+1)}}+\frac{1}{z^{2}}\sum\limits^{d}_{i=1}dx_{i}^{2}+g_{tt}dt^{2},
\end{equation}
where we have introduced $z\equiv\epsilon e^{-u}$. The authors of~\cite{Shiba:2013jja} studied the case when the
entangling region is a strip and observed a volume law for the holographic entanglement
entropy (HEE) when the boundary separation length of the strip is small enough.

In this paper we would like to perform a more detailed analysis on the holographic entanglement
entropy (HEE) of non-local quantum field theories, whose dual gravity background
is of the form~(\ref{nlgsec1}). In particular, we will consider different types of entangling
regions and the finite temperature counterparts. Note that since $g_{tt}$ cannot be determined
by cMERA, we have to focus on concrete models of gravity to construct the black hole
solutions. Here we mainly consider $(d+2)-$dimensional Einstein-Maxwell-dilaton (EMD) theory, which
admits metrics of the form~(\ref{nlgsec1}) as exact solutions. The finite temperature solutions are
also obtained in the EMD theory. Furthermore, if the Maxwell field is turned off, metrics of the form~(\ref{nlgsec1})
are still exact solutions of the resulting Einstein-dilaton theory.

After obtaining the exact solutions we turn to calculations of the HEE at zero temperature with entangling regions being a strip and a sphere.
We work out the results at next-to-leading order in the UV cutoff $\epsilon$ and find that the behavior of HEE depends on the value of $w$. We observe that in all the cases, the leading order of HEE exhibits a volume law as expected. Moreover, the next-to-leading order result shows a `milder' divergent behavior and may be independent of the cutoff when $w=d/2+1$. When $w=1$, the volume law always holds even at the next-to-leading order.

Generically it is very difficult to obtain analytic expressions for the HEE at finite temperature. However, we can obtain analytic results for a strip in both the high temperature limit and the low temperature limit, as claimed in~\cite{Fischler:2012ca}. In the low temperature limit, the extremal surface is restricted to be near the boundary region, hence the finite temperature corrections are small and can be computed perturbatively. In the high temperature limit, the extremal surface approaches the horizon and the leading contribution comes from the near-horizon region of the surface. We find that in the high temperature limit, the leading order result is always proportional to the volume of the entangling region, which is the same as observed in~\cite{Fischler:2012ca}. The low temperature results depend on the value of $w$. When $w\neq1$, we can work out the finite temperature corrections to the leading order and extract the finite contribution to the HEE at extremality by taking $T\rightarrow0$. However, when $w=1$, the finite temperature corrections are always of higher order in $\epsilon$, hence are not apparent at the leading order. The HEE is thus still given by the zero-temperature result.

The rest of the paper is organized as follows: In section 2 we construct the exact solutions in $(d+2)-$dimensional EMD theory, both at extremality and at finite temperature. We also discuss properties of similar solutions in Einstein-scalar theory. In section 3 we calculate the HEE in zero temperature background with strip and sphere entangling regions. The finite temperature corrections to HEE are studied in section 4 in the high temperature limit and low temperature limit respectively, where analytic expressions are obtained. A summary and discussion is given in section 5.

\section{The gravity duals}
In this section we construct exact solutions of the form~(\ref{nlgsec1}) in $(d+2)-$dimensional Einstein-Maxwell-dilaton (EMD) theory.
The EMD theory can be considered as the low energy effective theory that characterizes the IR physics of many condensed matter physics holographically,
which was extensively studied in~\cite{Charmousis:2010zz}. Moreover, the EMD theory admits the hyperscaling violating backgrounds as exact solutions, which exhibits the logarithmic violation of the area law for the entanglement entropy when the hyperscaling violating parameter is $d-1$~\cite{Ogawa:2011bz, Huijse:2011ef}.

Let us consider the following $(d+2)-$dimensional EMD theory,
\begin{equation}
S=\int d^{d+2}x\sqrt{-g}\left[R-\frac{1}{2}(\nabla\Phi)^{2}-V(\Phi)-\frac{1}{4}Z(\Phi)F_{\mu\nu}F^{\mu\nu}\right],
\end{equation}
where we have set $16\pi G_{N}=1$ for convenience. Here $V(\Phi)$ and $Z(\Phi)$ denote the scalar potential and the effective gauge coupling respectively.
The corresponding equations of motion are given by
\begin{equation}
\partial_{\mu}(\sqrt{-g}Z(\Phi)F^{\mu\nu})=0,
\end{equation}
\begin{equation}
\partial_{\mu}(\sqrt{-g}\partial^{\mu}\Phi)=\frac{1}{4}\sqrt{-g}\frac{\partial Z(\Phi)}{\partial\Phi}F_{\rho\sigma}F^{\rho\sigma}
+\sqrt{-g}\frac{\partial V}{\partial \Phi},
\end{equation}
\begin{eqnarray}
& &R_{\mu\nu}-\frac{1}{2}Rg_{\mu\nu}+\frac{1}{2}g_{\mu\nu}V(\Phi)-\frac{1}{2}\nabla_{\mu}\Phi\nabla_{\nu}\Phi\nonumber\\
& &+\frac{1}{4}g_{\mu\nu}(\nabla\Phi)^{2}-\frac{1}{2}Z(\Phi)F_{\mu\lambda}{F_{\nu}}^{\lambda}+\frac{1}{8}Z(\Phi)g_{\mu\nu}F_{\rho\sigma}
F^{\rho\sigma}=0.
\end{eqnarray}

To obtain exact solutions of the form~(\ref{nlgsec1}), we follow the recipe in~\cite{Ogawa:2011bz}, that is, we substitute the ansatz for the metric and matter fields into the equations of motion, from which we solve for the scalar potential $V(\Phi)$ and the effective gauge coupling $Z(\Phi)$. To ensure the physical sensibility of the solutions, we impose the Null Energy Condition (NEC). Consider the following ansatz
\begin{eqnarray}
ds^{2}_{d+2}&=&\frac{1}{z^{2}}[-f(z)dt^{2}+g(z)dz^{2}+\sum\limits^{d}_{i=1}dx_{i}^{2}],\nonumber\\
A&=&A_{t}(z)dt,~~~\Phi=\Phi(z),
\end{eqnarray}
where $f(z)$ and $g(z)$ are arbitrary functions of $z$. The NEC reads $T_{\mu\nu}N^{\mu}N^{\nu}\geq0$, where $N^{\mu}$ denotes any null vector and $T_{\mu\nu}$ is the energy momentum tensor, which is equal to the Einstein tensor $G_{\mu\nu}$. We can take the following components of the null vector,
\begin{equation}
N^{t}=\frac{1}{\sqrt{f(z)}},~~N^{z}=\frac{\cos\theta}{\sqrt{g(z)}},~~N^{x}=\sin\theta,
\end{equation}
where $\theta$ is an arbitrary constant. Then it can be seen that
\begin{eqnarray}
T_{\mu\nu}N^{\mu}N^{\nu}&=&-\frac{\sin^{2}\theta}{4zf(z)^{2}g(z)^{2}}[zg(z)f^{\prime2}(z)\nonumber\\
& &+f(z)\left(zf^{\prime}(z)g^{\prime}(z)+g(z)\left(2df^{\prime}(z)-2zf^{\prime\prime}(z)\right)\right)]\nonumber\\
& &-\cos^{2}\theta\frac{d(g(z)f^{\prime}(z)+g^{\prime}(z)f(z))}{2zf(z)g(z)^{2}}.
\end{eqnarray}
The NEC is satisfied if and only if
\begin{equation}
\label{nec1sec2}
g(z)f^{\prime}(z)+g^{\prime}(z)f(z)\leq0,
\end{equation}
\begin{equation}
\label{nec2sec2}
zg(z)f^{\prime2}(z)+f(z)[zf^{\prime}(z)g^{\prime}(z)+g(z)(2df^{\prime}(z)-2zf^{\prime\prime}(z))]\leq0,
\end{equation}

Comparing with the metric~(\ref{nlgsec1}), it is sufficient to choose
\begin{equation}
g(z)=\frac{z_{F}^{2}}{z^{2w}},~~~f(z)=\frac{1}{z^{2m}},
\end{equation}
where $z_{F}$ denotes certain energy scale at which the metric is a suitable description. Moreover, we impose $m\geq0, w\geq0$ so that the boundary of the spacetime is located at $z=0$. Then the NEC gives
\begin{equation}
\label{neciqsec2}
m+w\geq0,~~~d+1+m-w\geq0,
\end{equation}
Since we require $m\geq0, w\geq0$, the first inequality in~(\ref{neciqsec2}) holds automatically, while the second inequality leads to
\begin{equation}
w\leq m+d+1.
\end{equation}

Finally we obtain the solutions for $A_{t}(z), V(\Phi), Z(\Phi)$ and $\Phi(z)$,
\begin{eqnarray}
& &A_{t}^{\prime}=\frac{A_{0}}{Z(\Phi)}\sqrt{f(z)g(z)}z^{d-2},~~~V(\Phi)=-(m+d)(m+d+1-w)z^{2w}/z_{F}^{2},\nonumber\\
& &Z(\Phi)=\frac{A_{0}^{2}z_{F}^{2}z^{2d-2w}}{2m(m+d+1-w)},~~~\Phi^{\prime2}=\frac{2d(m+w)}{z^{2}},
\end{eqnarray}
where $A_{0}$ is an integration constant. It can be easily seen that such solutions can be reduced to various existing solutions in the literatures. For example, if we take $m=d/\eta, w=1$, we obtain the `semi-local' background which is conformal to $AdS_{2}\times\mathbb{R}^{d}$~\cite{Erdmenger:2013rca}.
When $m>0, w=0$, the solutions become those studied in~\cite{Taylor:2008tg} with Lifshitz scaling. If $m=0,w=0$, we arrive at pure AdS.

The EMD theory also admits the following finite temperature counterparts
\begin{eqnarray}
ds^{2}_{d+2}&=&\frac{1}{z^{2}}[-\frac{h(z)}{z^{2m}}dt^{2}+\frac{z_{F}^{2}}{z^{2w}h(z)}dz^{2}+\sum\limits^{d}_{i=1}dx_{i}^{2}],\nonumber\\
h(z)&=&1-\left(\frac{z}{z_{H}}\right)^{\gamma},~~~\gamma\equiv m+d+1-w,
\end{eqnarray}
while the gauge field and dilaton remain the same as the zero temperature solutions. The corresponding temperature and entropy density are given by
\begin{equation}
T=\frac{\gamma}{4\pi}z_{H}^{w-m-1},~~~s=4\pi z_{H}^{-d},
\end{equation}
which leads to
\begin{equation}
s\propto T^{\frac{d}{m+1-w}}.
\end{equation}
If we require that the entropy density vanishes in the $T\rightarrow0$ limit, there exists another condition for the parameters,
\begin{equation}
\label{constraint2sec2}
m+1-w>0.
\end{equation}
We will take~(\ref{constraint2sec2}) as an additional constraint in the following analysis. Note also that once~(\ref{constraint2sec2}) is imposed, the specific heat is positive.

It should be pointed out that such backgrounds, including both the extremal and finite temperature ones, are also exact solutions of Einstein-dilaton theory
\begin{equation}
S=\int d^{d+2}x\sqrt{-g}[R-\frac{1}{2}(\nabla\Phi)^{2}-V(\Phi)].
\end{equation}
The corresponding solutions can be obtained by simply setting $m=0$ and the Maxwell field is decoupled from the EMD theory in this limit. The properties of such black holes have been analysed in~\cite{Charmousis:2010zz}. It can be seen that when $w=1$, the temperature is constant. Furthermore, the entropy density obeys
\begin{equation}
s\propto T^{\frac{d}{1-w}},
\end{equation}
thus $s$ diverges as $T\rightarrow0$ when $w>1$. For a more detailed analysis on the phase diagram, see section 2.5 of~\cite{Charmousis:2010zz}. We will consider the $m>0$ cases in the following with $w<d$.
\section{Volume-law behavior of the HEE}
In this section we study the holographic entanglement entropy (HEE) in the zero temperature background, with the entangling regions being both a strip and a sphere. The HEE in similar backgrounds has been studied in~\cite{Liu:2013una, Erdmenger:2013rca, Alishahiha:2012ad, Kulaxizi:2012gy}, where~\cite{Erdmenger:2013rca} and~\cite{Kulaxizi:2012gy} focused on the $w=1$ case while $w\geq1$ and $w<1$ cases were extensively studied in~\cite{Liu:2013una}. However, in the above mentioned papers the background~(\ref{nlgsec1}) was considered as the IR metric and the asymptotic UV geometry was still AdS. Here we treat our exact solutions as the full metric, aiming at extracting the non-local nature of HEE.

We concentrate on the cases with $w\geq1$ in the following, as the perturbative expansion of the HEE breaks down for $w<1$. As a result, $w$ should take values in the following parameter range,
\begin{equation}
1\leq w<m+1,
\end{equation}
where we have applied~(\ref{constraint2sec2}). Therefore if we insist that the entropy density vanishes in the extremal limit, $m$ should be positive and the black hole solution in Einstein-dilaton theory should not be taken into account.
We find that when the size of the entangling region is sufficiently small, the leading order behavior of the HEE exhibits a volume law as expected. The `small size' limit allows us to work perturbatively in terms of the UV cutoff $\epsilon$. We obtain the HEE to the next-to-leading order in $\epsilon$ and find that it always gives a `milder' divergent term when $w>1$. In particular, when $w=d/2+1$, the next-to-leading order term is finite and independent of $\epsilon$. However, when $w=1$, the volume law behavior still holds at the next-to-leading order. We will set $z_{F}=1$ in the following sections.
\subsection{The strip}
\label{sec31}
Let us consider the simplest case where the entangling region is a strip,
$$x\equiv x_{1}\in[-l/2, l/2],~~ x_{i}\in[0, L], ~~i=2,\cdots,d,$$
with $L\rightarrow\infty$. The induced metric is given by
\begin{equation}
ds^{2}_{\rm ind}=\frac{1}{z^{2}}\left(1+\frac{\dot{z}^{2}}{z^{2w}}\right)dx^{2}+\frac{1}{z^{2}}\sum\limits^{d}_{i=2}dx_{i}^{2},
\end{equation}
where we have parameterized $z=z(x)$ and $\dot{z}=dz/dx$.
The minimal surface area reads
\begin{eqnarray}
\label{areasec31}
A&=&\int d^{d}x\mathcal{L}\nonumber\\
&=&2L^{d-1}\int^{\frac{l}{2}}_{0}dx\frac{1}{z^{d}}\sqrt{1+\frac{\dot{z}^{2}}{z^{2w}}}.
\end{eqnarray}
For the strip case, we impose the following boundary conditions
\begin{equation}
z(x=l/2)=z_{\epsilon}=\epsilon,~~~z(x=0)=z_{\ast},
\end{equation}
where $\epsilon$ denotes the UV cutoff and $z_{\ast}$ is the turning point of the extremal surface, that is, the point of closest approach of the extremal surface into the bulk spacetime.

Since the Lagrangian $\mathcal{L}$ does not explicitly contain $z$, there exists a conserved quantity and the corresponding Hamiltonian is given by
\begin{equation}
H=\dot{z}\frac{\partial\mathcal{L}}{\partial\dot{z}}-\mathcal{L}=-\frac{1}{z^{d}\sqrt{1+\frac{\dot{z}^{2}}{z^{2w}}}}.
\end{equation}
Note that at the turning point $z=z_{\ast}, \dot{z}=0$, which gives $H=-1/z_{\ast}^{d}$, therefore we can solve for $\dot{z}$,
\begin{equation}
\label{dotzsec31}
\dot{z}=z^{w}\sqrt{\frac{z_{\ast}^{2d}}{z^{2d}}-1}.
\end{equation}
Furthermore, when $l$ is sufficiently small, which means that the extremal surface is close to the boundary, we have the following expansion
for $z_{\epsilon}$
\begin{equation}
\label{zezasec31}
z_{\epsilon}=z_{\ast}+\frac{1}{8}\ddot{z}(z_{\ast})l^{2}=z_{\ast}-\frac{d}{8}z_{\ast}^{2w-1}l^{2},
\end{equation}
where we have used~(\ref{dotzsec31}) to evaluate $\ddot{z}(z_{\ast})$.

By plugging~(\ref{dotzsec31}) into~(\ref{areasec31}), the minimal surface area can be rewritten as
\begin{equation}
\label{Asec31}
A=2L^{d-1}\int^{z_{\ast}}_{z_{\epsilon}}dz\frac{z_{\ast}^{d}}{z^{w+d}\sqrt{z_{\ast}^{2d}-z^{2d}}}.
\end{equation}
Finally by combining~(\ref{zezasec31}) and~(\ref{Asec31}), we arrive at the holographic entanglement entropy
\begin{equation}
 S=4\pi A=2\pi\frac{lL^{d-1}}{\epsilon^{d}}\left(1-\frac{d^{2}}{8}l^{2}\epsilon^{2w-2}\right).
 \end{equation}
 Some remarks are in order. First of all, it can be easily seen that the leading term obeys a volume law as expected. Secondly, when $w=1$, the volume law still holds at the next-to-leading order. Thirdly, the next-to-leading order contribution always shows a `milder' divergent behavior when $w>1$. In particular,
 when $w=d/2+1$, the next-to-leading order term is independent of $\epsilon$. Since we have imposed $w<d$ at the end of section 2, the existence of such a universal term requires $d>2$, which is consistent. It indicates that we may obtain finite contribution to the HEE, which is proportional to $l^{3}L^{d-1}$. This will be confirmed in the low temperature limit expansion in the next section. Finally, if $w<1$, the second term diverges more quickly than the leading term, which suggests that perturbative expansion of the HEE breaks down when $w<1$. Thus the results in pure AdS~\cite{Ryu:2006ef} cannot be reproduced by simply taking $m=w=0$.

 It was pointed out in~\cite{Shiba:2013jja} that the volume law holds only when $l$ is sufficiently small. When $l$ is large enough,
 the two disconnected minimal surfaces dominate and the corresponding HEE is proportional to $L^{d-1}$. This example of `phase transition' has been observed in previous investigations, e.g. in~\cite{Erdmenger:2013rca, Kulaxizi:2012gy, Klebanov:2007ws}.

\subsection{The sphere}
In this subsection we consider the case where the entangling region is a sphere, in order to see if the volume-law behavior is universal.
Generically there exist two possible topologies of the extremal surface, i.e. cylinder and disk, with a sphere entangling region. As analyzed in~\cite{Liu:2013una}, in the large $l$ limit, when $w>1$ only the cylinder topology is possible and when $w=1$ the disk topology is possible.
However, since we are focusing on the small $l$ limit and expecting to see the volume law behavior, we will discuss the disk case in the following.

We parameterize the spatial part of the metric as follows
\begin{equation}
\sum\limits^{d}_{i=1}dx_{i}^{2}=d\rho^{2}+\rho^{2}d\Omega^{2}_{d-1},
\end{equation}
where $\rho\in[0, l]$ and $d\Omega^{2}_{d-1}$ denotes the metric on a unit sphere.
The induced metric is given by
\begin{equation}
ds^{2}_{\rm ind}=\frac{1}{z^{2}}\left(1+\frac{\dot{z}^{2}}{z^{2w}}\right)d\rho^{2}+\frac{\rho^{2}}{z^{2}}d\Omega^{2}_{d-1},
\end{equation}
where we have parameterized $z=z(\rho)$ and $\dot{z}=dz/d\rho$.
We can also obtain the minimal surface area
\begin{equation}
A=\int d\Omega_{d-1}d\rho\frac{\rho^{d-1}}{z^{d}}\sqrt{1+\frac{\dot{z}^{2}}{z^{2w}}},
\end{equation}

Unlike the strip case, there is no Hamiltonian for the current case, but we can still derive the equation of motion
\begin{equation}
\partial_{\rho}\left(\frac{\rho^{d-1}}{z^{d}}\frac{\dot{z}^{2}/z^{2w}}{\sqrt{1+\frac{\dot{z}^{2}}{z^{2w}}}}\right)
=-\frac{d\rho^{d-1}}{z^{d+1}}\sqrt{1+\frac{\dot{z}^{2}}{z^{2w}}}-\frac{w\rho^{d-1}}{z^{2w+d+1}}\frac{\dot{z}^{2}}{\sqrt{1+\frac{\dot{z}^{2}}{z^{2w}}}},
\end{equation}
with boundary conditions
\begin{equation}
z(\rho=l)=\epsilon,~~~z(\rho=0)=z_{\ast}.
\end{equation}
Note also that $\dot{z}(\rho=0)=0$ at the turning point $z=z_{\ast}$.
The subsequent calculations are similar to the previous subsection for the strip case. In the
small $l$ limit we can expand $z_{\ast}$ as
\begin{equation}
z_{\ast}=\epsilon+\frac{d}{2}z^{2w-1}_{\ast}l^{2}.
\end{equation}
Therefore the holographic entanglement entropy reads,
\begin{equation}
S=4\pi A=\frac{4\pi}{d}\frac{l^{d}\Omega_{d-1}}{\epsilon^{d}}\left(1-\frac{d}{2}\epsilon^{2w-2}l^{2}\right),
\end{equation}
 Once again we can see that the leading term obeys a volume law as expected and the volume law still holds at the next-to-leading order at $w=1$.
 when $w=d/2+1$, the next-to-leading order term is independent of $\epsilon$. The perturbative expansion of the HEE breaks down when $w<1$ and the results in pure AdS~\cite{Ryu:2006ef} cannot be reproduced by simply taking $m=w=0$. In sum, the features of HEE with a spherical entangling region are qualitatively similar to those of the strip case.

\begin{figure}
\begin{center}
\vspace{-1cm}
\hspace{-0.5cm}
\includegraphics[angle=0,width=0.45\textwidth]{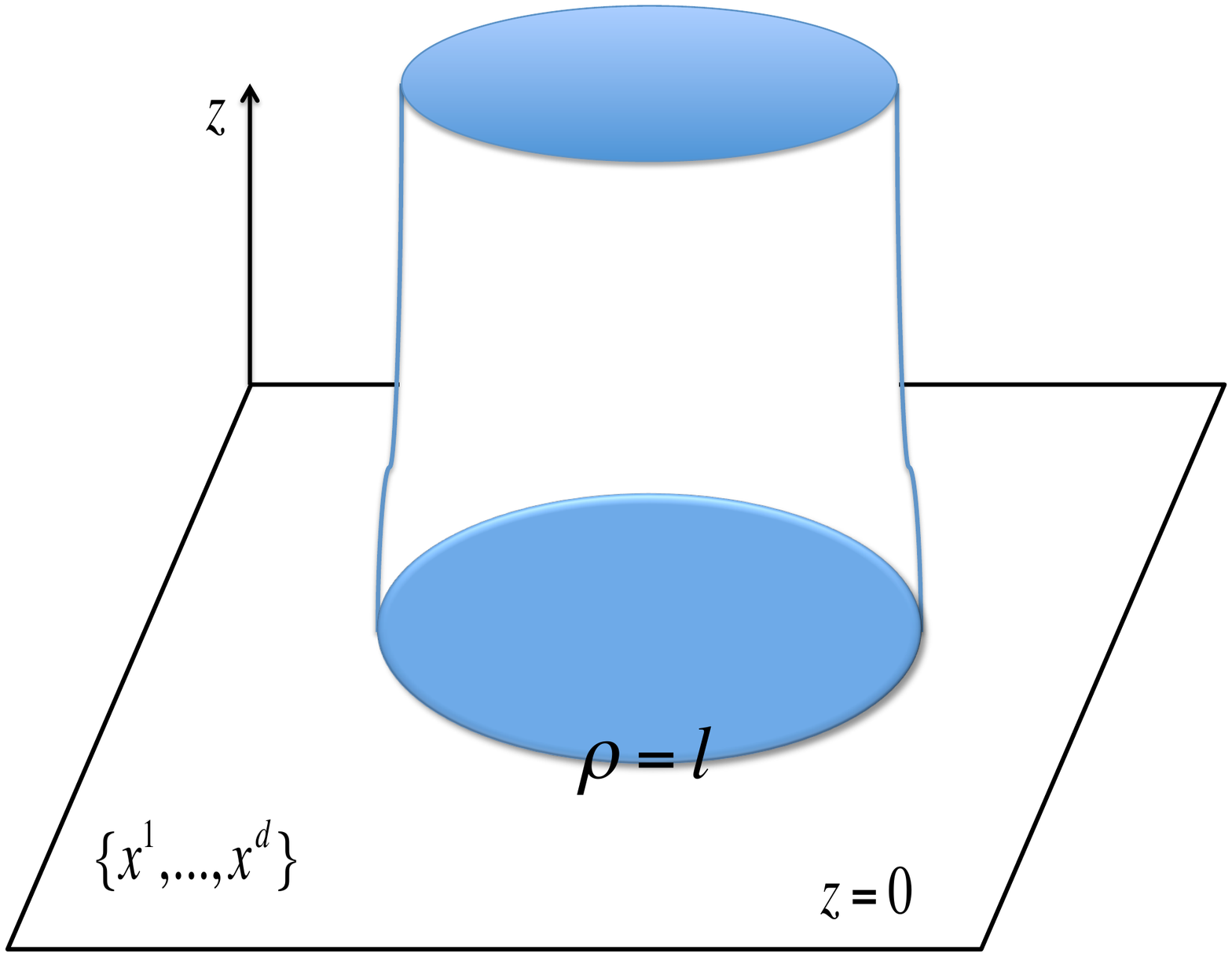}
\includegraphics[angle=0,width=0.45\textwidth]{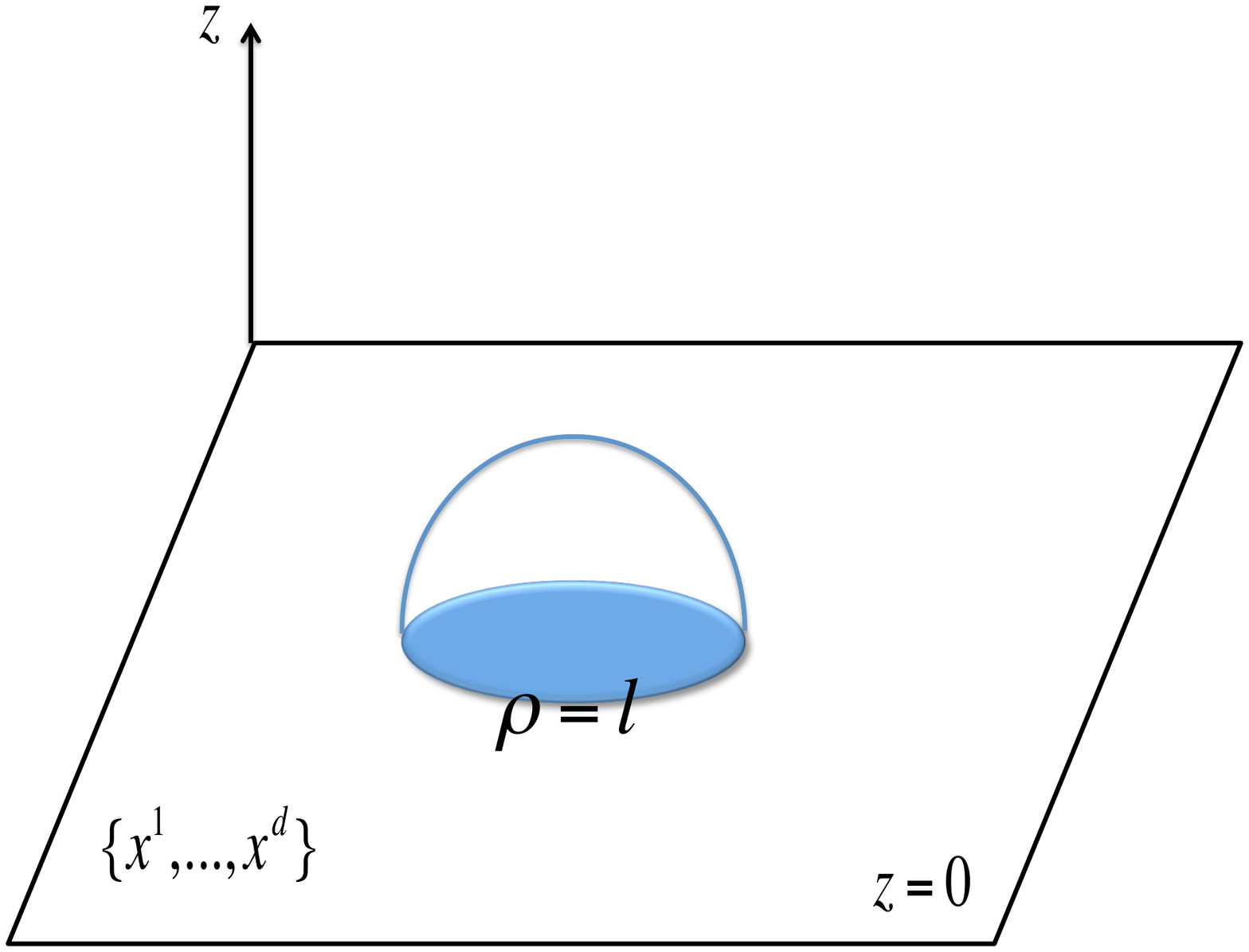}
\caption{\small The minimal surfaces with different topologies at $w>1$. Left: The minimal surface with cylinder topology when $l$ is large. Right: The minimal surface
with disk topology when $l$ is sufficiently small.}
\label{gamma008}
\end{center}
\end{figure}

 It has been pointed out in~\cite{Liu:2013una} that when $w>1$, only the minimal surface with cylinder topology is possible in the large $l$ limit, hence the HEE is proportional to $Ll^{d-2}\Omega_{d-2}$, which is an area law with $L$ being the `height' of the cylinder. It suggests that in this case, we also have a `phase transition' similar to the strip case and the HEE scales from a volume law to an area law. Minimal surfaces with different topologies at $w>1$ are shown in~\ref{gamma008}. However, when $w=1$, the minimal surface with disk topology still dominates~\cite{Liu:2013una}, but the HEE obeys an area law~\cite{Erdmenger:2013rca}. Therefore there is no `phase transition' for $w=1$ but the HEE still undergoes a `volume/area law' transition.

\section{HEE at finite temperature}
In this section we calculate the HEE for non-local field theories at finite temperature. Generically, it is very difficult to obtain analytic expressions for the HEE at finite temperature and one has to resort to numerics. However, it has been observed in~\cite{Fischler:2012ca} that it is possible to obtain analytic expressions for the HEE in both the high temperature limit and the low temperature limit. In the low temperature limit, the extremal surface is restricted to be near the boundary region, hence the finite temperature corrections are small and can be computed perturbatively. In the high temperature limit, the extremal surface approaches the horizon and the leading contribution comes from the near-horizon region of the surface. For simplicity we just consider the strip entangling region and discuss the $w\neq1$ and $w=1$ cases separately.

In this section we paramaterize $x=x(z)$, which results in the following induced metric
\begin{equation}
ds^{2}_{\rm ind}=\frac{1}{z^{2}}\left[\left(\frac{1}{z^{2w}h(z)}+x^{\prime^2}\right)dz^{2}+\sum\limits^{d}_{i=2}dx_{i}^{2}\right],
\end{equation}
where $x^{\prime}=dx/dz$.
The corresponding minimal surface area reads
\begin{equation}
A=\int d^{d-1}xdz\mathcal{L}=L^{d-1}\int dz\frac{1}{z^{d}}\sqrt{x^{\prime2}+\frac{1}{z^{2w}h(z)}}.
\end{equation}
Similarly, there exists a conserved quantity
\begin{equation}
\frac{\partial\mathcal{L}}{\partial x^{\prime}}=\frac{1}{z^{d}}\frac{x^{\prime}}{\sqrt{x^{\prime2}+\frac{1}{z^{2w}h(z)}}}.
\end{equation}
Note that at the turning point $z=z_{\ast}$, $x^{\prime}$ diverges, which gives
\begin{equation}
x^{\prime}=\frac{z^{d-w}}{z_{\ast}^{d}}h(z)^{-1/2}\left(1-\frac{z^{2d}}{z_{\ast}^{2d}}\right)^{-1/2}.
\end{equation}
The above expression enables us to express the boundary separation length as
\begin{equation}
l=2\int^{z_{\ast}}_{0}dz\frac{z^{d-w}}{z_{\ast}^{d}}h(z)^{-1/2}\left(1-\frac{z^{2d}}{z_{\ast}^{2d}}\right)^{-1/2}.
\end{equation}

Generally, the HEE can be divided into the divergent part and the finite part,
\begin{equation}
S_{A}=S_{\rm div}+S_{\rm finite}.
\end{equation}
When $l$ is small enough, the minimal surface is `localized' near the boundary and the divergent part of the HEE has been discussed in the previous section.
Note that $S_{\rm div}$ may exhibit a volume law or an area law, depending on the size of $l$. In the high temperature limit, the divergent part is given by
\begin{equation}
S_{\rm div}=\frac{8\pi L^{d-1}}{d+w-1}\frac{1}{\epsilon^{d+w-1}}.
\end{equation}
In the subsequent calculations we only focus on the finite part of the HEE, which can be evaluated from
the finite part of the minimal surface area
\begin{equation}
A_{\rm finite}=2L^{d-1}\int^{z_{\ast}}_{0}\frac{1}{z^{d+w}}h(z)^{-1/2}\left(1-\frac{z^{2d}}{z_{\ast}^{2d}}\right)^{-1/2},
\end{equation}
\subsection{$w\neq1$}
When $w\neq1$, we can introduce $u=z/z_{\ast}$, following~\cite{Fischler:2012ca}. Hence the boundary separation length can be rewritten as
\begin{equation}
l=2z_{\ast}^{1-w}\sum\limits^{\infty}_{n=0}\frac{\Gamma(n+\frac{1}{2})}{\sqrt{\pi}\Gamma(n+1)}\int^{1}_{0}duu^{n\gamma+d-w}(1-u^{2d})^{-1/2}
\left(\frac{z_{\ast}}{z_{H}}\right)^{n\gamma}.
\end{equation}
Let us assume that $w<d$ so that the integral at $n=0$ converges, which gives
\begin{equation}
\label{lwneq1}
l=2z_{\ast}^{1-w}\sum\limits^{\infty}_{n=0}\frac{\Gamma(n+\frac{1}{2})}{\Gamma(n+1)}\frac{\Gamma(\frac{1}{2}+\frac{k_{n}}{2d})}{k_{n}\Gamma(\frac{k_{n}}{2d})}
\left(\frac{z_{\ast}}{z_{H}}\right)^{n\gamma},
\end{equation}
where $k_{n}=n\gamma-w+1$. Note that in the large $n$ limit, the terms in the above series approximate to $\sim\frac{1}{n}(z_{\ast}{z_{H}})^{n\gamma}$, which does not converge as $z_{\ast}\rightarrow z_{H}$. Therefore we have to isolate the divergent terms,
\begin{eqnarray}
l&=&2z_{\ast}^{1-w}\frac{\sqrt{\pi}\Gamma(\frac{1}{2}+\frac{1-w}{2d})}{(1-w)\Gamma(\frac{1-w}{2d})}-\frac{\sqrt{2}}{\sqrt{\gamma d}}z_{\ast}^{1-w}\log\left[1-\left(\frac{z_{\ast}}{z_{H}}\right)^{\gamma}\right]\nonumber\\
& &+2z_{\ast}^{1-w}\sum\limits^{\infty}_{n=1}\left[
\frac{\Gamma(n+\frac{1}{2})}{\Gamma(n+1)}\frac{\Gamma(\frac{1}{2}+\frac{k_{n}}{2d})}{k_{n}\Gamma(\frac{k_{n}}{2d})}-\frac{1}{2d\gamma}\frac{1}{n}\right]
\left(\frac{z_{\ast}}{z_{H}}\right)^{n\gamma},
\end{eqnarray}
which ensures that the summation in~(\ref{lwneq1}) is convergent.

In the high temperature limit, the extremal surface approaches very close to the horizon, hence we can expand $z_{\ast}=z_{H}(1-\varepsilon)$,
where $\varepsilon$ is a very small quantity. In order to fix the value of $\varepsilon$, we can expand~(\ref{lwneq1}) to leading order in $\varepsilon$,
\begin{eqnarray}
\frac{\sqrt{2}}{\sqrt{\gamma d}}\log(\varepsilon\gamma)&=&-lz_{H}^{w-1}+\frac{2\sqrt{\pi}\Gamma(\frac{1}{2}+\frac{1-w}{2d})}{(1-w)\Gamma(\frac{1-w}{2d})}\nonumber\\
& &+2\sum\limits^{\infty}_{n=1}\left[\frac{\Gamma(n+\frac{1}{2})}{\Gamma(n+1)}\frac{\Gamma(\frac{1}{2}+\frac{k_{n}}{2d})}{k_{n}\Gamma(\frac{k_{n}}{2d})}
-\frac{1}{\sqrt{2\gamma d}n}\right],
\end{eqnarray}
which gives
\begin{equation}
\varepsilon=\mathcal{E}_{\rm ent}e^{-\sqrt{\gamma d/2}lz_{H}^{w-1}},
\end{equation}
where
\begin{equation}
\mathcal{E}_{\rm ent}=\frac{1}{\gamma}\exp\left\{\sqrt{\frac{\gamma d}{2}}\left[\frac{2\sqrt{\pi}\Gamma(\frac{1}{2}+\frac{1-w}{2d})}{(1-w)\Gamma(\frac{1-w}{2d})}
+2\sum\limits^{\infty}_{n=1}\left(\frac{\Gamma(n+\frac{1}{2})}{\Gamma(n+1)}\frac{\Gamma(\frac{1}{2}+\frac{k_{n}}{2d})}{k_{n}\Gamma(\frac{k_{n}}{2d})}
-\frac{1}{\sqrt{2\gamma d}n}\right)
\right]\right\}.
\end{equation}

We can also expand the finite part of the minimal surface area in a similar way,
\begin{eqnarray}
A_{\rm finite}&=&\frac{2L^{d-1}}{z_{\ast}^{d+w-1}}\int du\frac{1}{u^{d+w}\sqrt{1-u^{2d}}}\left(1-\left(\frac{z_{\ast}}{z_{H}}\right)^{\gamma}u^{\gamma}\right)^{-1/2}\nonumber\\
&=&\frac{2L^{d-1}}{z_{\ast}^{d+w-1}}\sum\limits^{\infty}_{n=0}\frac{\Gamma(n+\frac{1}{2})}{\Gamma(n+1)}\frac{\Gamma(\frac{1}{2}+\frac{k_{n}}{2d})}
{(k_{n}-d)\Gamma(\frac{k_{n}}{2d})}\left(\frac{z_{\ast}}{z_{H}}\right)^{n\gamma}.
\end{eqnarray}
After substituting the expression for $l$ given in~(\ref{lwneq1}), we arrive at
\begin{eqnarray}
A_{\rm finite}&=&\frac{2L^{d-1}}{z_{\ast}^{d+w-1}}\Big[\frac{l}{2}z_{\ast}^{w-1}+\frac{\sqrt{\pi}}{2(1-w)}\frac{\Gamma(\frac{1-w-d}{2d})}
{\Gamma(\frac{1-w}{2d})}\nonumber\\
& &+\sum\limits^{\infty}_{n=1}\frac{d}{k_{n}(k_{n}-d)}\frac{\Gamma(n+\frac{1}{2})}{\Gamma(n+1)}
\frac{\Gamma(\frac{1}{2}+\frac{k_{n}}{2d})}{\Gamma(\frac{k_{n}}{2d})}\left(\frac{z_{\ast}}{z_{H}}\right)^{n\gamma}\Big].
\end{eqnarray}
Note that when $n\rightarrow\infty$, the infinite series behaves as $\frac{1}{n^{2}}(z_{\ast}/z_{H})^{n\gamma}$ and the summation converges as $z_{\ast}\rightarrow z_{H}$. So it would be straightforward to take the limit $z_{\ast}\rightarrow z_{H}$ which gives
\begin{equation}
A_{\rm finite}=\frac{lL^{d-1}}{z_{H}^{d}}\left(1+\frac{1}{lz_{H}^{w-1}}\mathcal{S}_{\rm high}\right),
\end{equation}
where
\begin{equation}
\mathcal{S}_{\rm high}=2\left[\frac{\sqrt{\pi}}{2(1-w)}\frac{\Gamma(\frac{1-w-d}{2d})}
{\Gamma(\frac{1-w}{2d})}
+\sum\limits^{\infty}_{n=1}\frac{d}{k_{n}(k_{n}-d)}\frac{\Gamma(n+\frac{1}{2})}{\Gamma(n+1)}
\frac{\Gamma(\frac{1}{2}+\frac{k_{n}}{2d})}{\Gamma(\frac{k_{n}}{2d})}\right].
\end{equation}
Finally the finite part of the holographic entanglement entropy is given by
\begin{equation}
S_{\rm finite}=4\pi V\left(\frac{4\pi T}{\gamma}\right)^{\frac{d}{m+1-w}}\left(1+\frac{1}{l}\left(\frac{4\pi T}{\gamma}\right)^{\frac{w-1}{m+1-w}}
\mathcal{S}_{\rm high}\right),
\end{equation}
where $V=lL^{d-1}$. It can be seen that in the high temperature limit, the leading order finite-temperature contribution is proportional to the volume of the entangling region, which is the same as that observed in~\cite{Fischler:2012ca}. This is consistent with the fact that in the high temperature limit, the entanglement entropy approaches the thermal entropy. When $m=w=0$, we can reproduce the results obtained in~\cite{Fischler:2012ca}.

In the low temperature limit $l/z_{H}\ll1$ we can work out the finite-temperature contributions perturbatively,
\begin{equation}
z_{\ast}=z_{\ast0}\left(1-\lambda\left(\frac{l}{z_{H}}\right)^{\gamma}\right),
\end{equation}
where $\lambda$ is another very small quantity. This enables us to rewrite $l$ as follows,
\begin{equation}
l=2z_{\ast0}^{1-w}\frac{\sqrt{\pi}}{(1-w)}\frac{\Gamma(\frac{1-w-d}{2d})}
{\Gamma(\frac{1-w}{2d})}.
\end{equation}
Working to the leading order in $\lambda$, we obtain
\begin{equation}
\lambda=\frac{1}{2k_{1}}\frac{\Gamma(\frac{1}{2}+\frac{k_{1}}{2d})}{\Gamma(\frac{k_{1}}{2d})}
\frac{\Gamma(\frac{1-w}{2d})}{\Gamma(\frac{1}{2}+\frac{1-w}{2d})}\left(\frac{1-w}{2\sqrt{\pi}}\frac{\Gamma(\frac{1-w}{2d})}{\Gamma(\frac{1}{2}+
\frac{1-w}{2d})}\right)^{\frac{\gamma}{1-w}}l^{\frac{\gamma w}{1-w}}.
\end{equation}
Therefore the finite part of the minimal surface area reads
\begin{equation}
A_{\rm finite}=\frac{L^{d-1}}{l^{d-1}}l^{\frac{wd}{w-1}}\mathcal{S}_{0}\left(1+\mathcal{S}_{1}\left(\frac{l}{z_{H}}\right)^{\gamma}l^{\frac{\gamma w}{1-w}}\right),
\end{equation}
where
\begin{eqnarray}
& &\mathcal{S}_{0}=\frac{\sqrt{\pi}}{d}\frac{\Gamma(\frac{1-d-w}{2d})}{\Gamma(\frac{1-w}{2d})}
\left(\frac{2\sqrt{\pi}\Gamma(\frac{1}{2}+\frac{1-w}{2d})}{(1-w)\Gamma(\frac{1-w}{2d})}\right)^{\frac{d+w-1}{1-w}},\\
& &\mathcal{S}_{1}=\frac{2d}{\sqrt{\pi}}\frac{\Gamma(\frac{1-w}{2d})}{\Gamma(\frac{1-wd}{2d})}\lambda_{1},\\
& &\lambda_{1}=\lambda(d+w-1)\frac{\sqrt{\pi}\Gamma(\frac{1-d-w}{2d})}{2d\Gamma(\frac{1-w}{2d})}+\frac{\sqrt{\pi}\Gamma(\frac{k_{1}}{2d}-\frac{1}{2})}
{4d\Gamma(\frac{k_{1}}{2d})}.
\end{eqnarray}
Finally the finite part of the holographic entanglement entropy is given by
\begin{equation}
\label{sfinitewneq1}
S_{\rm finite}=4\pi\frac{L^{d-1}}{l^{d-1}}l^{\frac{wd}{w-1}}\mathcal{S}_{0}\left(1+\mathcal{S}_{1}l^{\frac{\gamma}{1-w}}
\left(\frac{4\pi T}{\gamma}\right)^{\frac{\gamma}{m+1-w}}\right).
\end{equation}

It should be pointed out that we can extract the finite contribution to the HEE in the extremal background.
If we take $T\rightarrow0$ in~(\ref{sfinitewneq1}), we can obtain
\begin{equation}
S_{\rm finite}=4\pi\frac{L^{d-1}}{l^{d-1}}l^{\frac{wd}{w-1}}\mathcal{S}_{0}.
\end{equation}
As a check of consistency, let us take $w=d/2+1$, which leads to $S_{\rm finite}\sim L^{d-1}l^{3}$. This agrees with the result obtained in section~\ref{sec31}.
\subsection{$w=1$}
The case $w=1$ should be treated separately, as some of the results presented in this subsection cannot be obtained by directly taking $w=1$ from the corresponding ones in the previous subsection. First of all, the boundary separation length $l$ can be expanded as follows
\begin{equation}
l=\frac{\pi}{d}+2\sum\limits^{\infty}_{n=1}\frac{\Gamma(n+\frac{1}{2})\Gamma(\frac{1}{2}+\frac{n\gamma_{1}}{2d})}
{n\gamma_{1}\Gamma(n+1)\Gamma(\frac{n\gamma_{1}}{2d})}\left(\frac{z_{\ast}}{z_{H}}\right)^{n\gamma_{1}},
\end{equation}
where $\gamma_{1}=m+d$. Note that the leading constant term has been obtained in the zero temperature background in~\cite{Erdmenger:2013rca} and the summation in the series starts from $n=1$ because the $n=0$ term goes to zero in this case. Similarly we have to isolate the singular term
\begin{eqnarray}
l&=&\frac{\pi}{d}-\frac{\sqrt{2}}{\sqrt{\gamma_{1}d}}\log\left[1-\left(\frac{z_{\ast}}{z_{H}}\right)^{\gamma_{1}}\right]\nonumber\\
& &+2\sum\limits^{\infty}_{n=1}\left[\frac{\Gamma(n+\frac{1}{2})\Gamma(\frac{1}{2}+\frac{n\gamma_{1}}{2d})}
{n\gamma_{1}\Gamma(n+1)\Gamma(\frac{n\gamma_{1}}{2d})}-\frac{1}{\sqrt{2d\gamma_{1}}}\frac{1}{n}\right]\left(\frac{z_{\ast}}{z_{H}}\right)^{n\gamma_{1}},
\end{eqnarray}
in order to obtain a convergent summation.

In the high temperature limit, we can expand $z_{\ast}=z_{H}(1-\varepsilon)$,
where the small quantity $\varepsilon$ at leading order is given by
\begin{equation}
\varepsilon=\mathcal{E}_{\rm ent1}e^{-\sqrt{\gamma_{1}d/2}l},
\end{equation}
with
\begin{equation}
\mathcal{E}_{\rm ent1}=\frac{1}{\gamma_{1}}\exp\left[\sqrt{\frac{d\gamma_{1}}{2}}\left(\frac{\pi}{d}+2\sum\limits^{\infty}_{n=1}
\left[\frac{\Gamma(n+\frac{1}{2})\Gamma(\frac{1}{2}+\frac{n\gamma_{1}}{2d})}
{n\gamma_{1}\Gamma(n+1)\Gamma(\frac{n\gamma_{1}}{2d})}-\frac{1}{\sqrt{2d\gamma_{1}}}\frac{1}{n}\right]\right)\right].
\end{equation}
Furthermore, the finite part of the minimal surface area can be rewritten as
\begin{equation}
A_{\rm finite}=\frac{2L^{d-1}}{z_{\ast}^{d}}\left(\frac{dl}{2\pi}+\sum\limits^{\infty}_{n=1}\frac{d}{n\gamma_{1}(n\gamma_{1}-d)}
\frac{\Gamma(n+\frac{1}{2})}{\Gamma(n+1)}\frac{\Gamma(\frac{n\gamma_{1}}{2d}+\frac{1}{2})}{\Gamma(\frac{n\gamma_{1}}{2d})}
\left(\frac{z_{\ast}}{z_{H}}\right)^{n\gamma_{1}}\right).
\end{equation}
Note that the summation converges as $z_{\ast}\rightarrow z_{H}$, so it is straightforward to take such a limit, which results in
\begin{equation}
A_{\rm finite}=lL^{d-1}\frac{d}{\pi z_{H}^{d}}+\frac{L^{d-1}}{z_{H}^{d}}\mathcal{S}_{\rm high1},
\end{equation}
where
\begin{equation}
\mathcal{S}_{\rm high1}=\sum\limits^{\infty}_{n=1}\frac{d}{n\gamma_{1}(n\gamma_{1}-d)}
\frac{\Gamma(n+\frac{1}{2})}{\Gamma(n+1)}\frac{\Gamma(\frac{n\gamma_{1}}{2d}+\frac{1}{2})}{\Gamma(\frac{n\gamma_{1}}{2d})}.
\end{equation}
Finally the finite part of the holographic entanglement entropy is given by
\begin{equation}
S_{\rm finite}=4dV\left(\frac{4\pi T}{\gamma_{1}}\right)^{\frac{d}{m}}+4\pi L^{d-1}\mathcal{S}_{\rm high1}
\left(\frac{4\pi T}{\gamma_{1}}\right)^{\frac{d}{m}},
\end{equation}
where $V=lL^{d-1}$. Here the leading order finite-temperature contribution is also proportional to the volume, which is the case as expected.
The second term, even though being proportional to the area, has the same power in $T$.

Note that in the zero temperature limit, the boundary separation length is always constant and the disconnected minimal surfaces dominate.
Therefore there is no turning point and the perturbative analysis in previous subsection may not work. This can also be seen directly by taking
$z_{\ast}\rightarrow\epsilon$ in $l$ and $S_{\rm finite}$, which leads to
\begin{equation}
l=\frac{\pi}{d}+\frac{\sqrt{\pi}}{\gamma_{1}}\frac{\Gamma(\frac{1}{2}+\frac{\gamma_{1}}{2d})}{\Gamma(\frac{\gamma_{1}}{2d})}
\frac{\epsilon^{\gamma_{1}}}{z_{H}^{\gamma_{1}}},
\end{equation}
\begin{equation}
S_{\rm finite}=4\pi A_{\rm finite}=4\pi\mathcal{S}_{\gamma_{1}}\epsilon^{\gamma_{1}}\left(\frac{4\pi T}{\gamma_{1}}\right)^{\frac{\gamma_{1}}{m}},
\end{equation}
where
\begin{equation}
\mathcal{S}_{\gamma_{1}}=\frac{\sqrt{\pi}}{4d}\frac{\Gamma(\frac{\gamma_{1}}{2d}-\frac{1}{2})}{\Gamma(\frac{\gamma_{1}}{2d})}.
\end{equation}
Therefore the finite temperature corrections only appear at order $O(\epsilon^{\gamma_{1}})$ with $\gamma_{1}=m+d$. It suggests that for the $w=1$ case in the low temperature limit, the finite-temperature contributions do not appear at the leading order.
\section{Summary and discussion}
It has been well established that in a local quantum field theory with a UV fixed point, the entanglement entropy obeys the area law. This universal behavior has been confirmed by the AdS/CFT correspondence. However, there are examples in quantum many-body physics where the entanglement entropy exhibits a volume law. The volume-law behavior may be closely related to the non-locality of the corresponding field theory. Therefore it would be very interesting to explore the volume-law behavior in the context of holography. In this paper we perform a detailed study on the holographic entanglement entropy of non-local field theories, whose dual metric was constructed in~\cite{Nozaki:2012zj} via cMERA. Since the metric component $g_{tt}$ cannot be determined by cMERA, we have to consider concrete effective gravity models to study the holographic entanglement entropy at finite temperature. We find that the metric~(\ref{nlgsec1}) as well as its finite temperature counterpart can emerge as exact solutions of $(d+2)-$dimensional Einstein-Maxwell-dilaton theory. We make sure the physical sensibility of such solutions by imposing the null energy condition and requiring that the entropy density vanishes in the zero temperature limit.

We study the HEE up to the next-to-leading order in the zero temperature background with entangling regions being a strip and a sphere. We find that when the size of the entangling region is small enough, though the next-to-leading order behavior depends on $w$, the leading order term in the HEE always exhibits a volume law. We also observe that when $w=d/2+1$, the next-to-leading order term is independent of the UV cutoff $\epsilon$. Furthermore, when $w=1$ the volume law still holds at the next-to-leading order. When the entangling region is a sphere, for $w>1$ the extremal surface with cylinder topology dominates over that with disk topology as $l$ increases and the resulting HEE obeys an area law, which is similar to the `phase transition' observed for the strip case where the disconnected surfaces dominate as $l$ grows. When $w=1$, the extremal surface with disk topology always dominates but the HEE also undergoes a `volume/area law' transition. We also investigate the finite-temperature corrections to the HEE analytically both in the high temperature limit and the low temperature limit. In the high temperature limit the leading order contribution is always proportional to the volume of the entangling region, which agrees with the fact that the entanglement entropy approaches the thermal entropy in this limit. In the low temperature limit the corrections can be evaluated pertubatively and we can extract the finite contribution to the HEE at extremality by taking $T\rightarrow0$. The resulting expression qualitatively agrees with the constant term in the zero temperature background when $w=d/2+1$. When $w=1$ the finite part of the HEE does not receive any corrections at the leading order.

Finally let us mention some future directions. One physical observable worth further investigation is the mutual information, which is defined as
\begin{equation}
I(A, B)=S_{A}+S_{B}-S_{A\cup B},
\end{equation}
for two disjoint subsystems $A$ and $B$. The mutual information is non-negative due to strong subadditivity. It has been known that for local field theories, the mutual information undergoes a phase transition if $A$ and $B$ have width $l$ and separation $x$~\cite{Headrick:2010zt},
\begin{eqnarray}
I(A,B)&=&2S(l)-S(x)-S(2l+x),~~~x/l\leq a_{c},\nonumber\\
I(A,B)&=&0,~~~x/l>a_{c},
\end{eqnarray}
where $a_{c}$ denotes the critical value of $x/l$. The holographic mutual information of non-communicative super Yang-Mills theory has been qualitatively analyzed in~\cite{Karczmarek:2013xxa}: When $l$ and $x$ are very small, the HEE obeys a volume law and hence $I(A,B)=0$. The phase transition may be observed when $l$ is large enough while $x$ is still very small. It is expected that similar behavior also holds in our background and it would be interesting to calculate the finite temperature contributions along the line of~\cite{Fischler:2012uv}.

When $w>1$, the HEE undergoes a phase transition, i.e. minimal surfaces of different types dominate as the size of the entangling region changes. Such kind of phase transitions were first observed in confining geometries in string theory~\cite{Nishioka:2006gr} and later were identified as a probe of confinement in~\cite{Klebanov:2007ws}. One may ask if there is potential connection between confinement and phase transitions of HEE. Recently this issue was studied in~\cite{Kol:2014nqa}, where the sufficient conditions for the phase transition of HEE are determined. Moreover, in the non-local QFT under consideration, the phase transition is absent unless a cutoff to the QFT is added or unless the QFT is UV completed. It would be interesting to perform parallel investigations on our background to further explore the relation between confinement and phase transition of HEE.

Note that our calculations are purely holographic and do not involve any field theory calculations. Therefore it would be very interesting to study the entanglement entropy at finite temperature through field theory analysis. For example, it may be plausible to reconstruct the metric at finite temperature via cMERA following~\cite{Mollabashi:2013lya} and study the corresponding quantum quench. It may also be interesting to generalize the analysis in~\cite{Karczmarek:2013jca} to field theories dual to our backgrounds. We leave these fascinating subjects to future work.

\bigskip \goodbreak \centerline{\bf Acknowledgments}
\noindent
DWP would like to thank the `HoloGrav' network of European Science Foundation for a short visit grant, and the
string theory group at University of Amsterdam for hospitality, where this work was initiated. DWP would also like to thank Carlos Nunez for helpful correspondence. This work was supported
by Alexander von Humboldt Foundation and a research grant from Max Planck Society.

\newpage

\end{document}